# SUPER-BRIDGES SUSPENDED OVER CARBON NANOTUBE CABLES

Alberto Carpinteri, Nicola M. Pugno
Dept. of Structural Engineering and Geotechnics
Politecnico di Torino, Corso Duca degli Abruzzi 24, 10129, Torino, ITALY
nicola.pugno@polito.it; Tel. +39 011 564 4902; Fax. +39 011 564 4899

## Abstract

In this paper the new concept of "super-bridges", i.e. kilometre-long bridges suspended over carbon nanotube cables, is introduced. The analysis shows that the use of realistic (thus defective) carbon nanotube bundles as suspension cables can enlarge the current limit main span by a factor of ~3. Too large compliance and dynamic self-excited resonances could be avoided by additional strands, rendering the super-bridge anchored as a spider cobweb. As an example, we have computed the limit main spans of the current existing 19 suspended-deck bridges longer than one kilometre assuming to have substituted their cables with carbon nanotube bundles (thus maintaining the same geometry, with the exception of the length) finding spans of up to ~6.3 Km. We thus suggest that the design of the Messina bridge in Italy, that would require a main span of ~3.3 Km, could benefit by the use of carbon nanotube bundles. We believe that their use is the sole feasible solution of the near future. The plausibility of these affirmations are confirmed by a statistical analysis on the existing 100 longest suspended bridges, which follow a Zipf's law with an exponent of 1.1615: we have found a Moore-like (i.e. exponential) law, in which the doubling of the capacity (here the main span) per year is substituted by the factor 1.0138. Such a law predicts the realization of the Messina bridge using conventional materials only around the half of the present century, whereas it could be expected in a nearest future if carbon nanotube bundles were used. A simple cost analysis concludes the paper.

## 1. Introduction

Suspension bridges are one of the oldest types of bridge. Early suspended bridges consisted of at least three cables made from vines, where people walked on the main rope to cross using the other two cables for self-balancing. Simple suspension bridges, with deck made from vines suspended on two cables, date back at least to 285BC (Peters, 1987) in China. Other similar suspended bridges are recorded in Tibet. Simple suspension bridges using iron chains are also documented in China and the Himalayas, although their earliest date is unclear, probably around the 15th century, perhaps built by Tibet monks (Peters, 1987). In our days a huge number of suspended bridges exists, e.g. 100 with spans longer than 325m, 19 of them reaching main spans longer than 1Km, see Table 1.

The main forces in a suspended-deck bridge are strong tension in the cables and compression in the pillars. The deck is, at least statically, poorly solicited. The key element is undoubtedly the main cable, which requires a material with a very high strength. The situation is totally different with that observable in stayed bridges, in which the strands are highly solicited and used to increase the stiffness, rather than the strength, of the bridge. In addition, the deck is extremely compressed by the action of the strands and is thus the element that limits the main span of the bridge. This shows that the use of super-strong strands is not very effective, suggesting that a suspended-deck rather than a stayed structure is required in order to increase the bridge main span by using super-strong cables.

Carbon nanotube bundles are ideal candidates for producing super-strong cables, thus kilometre-long suspended super-bridges, but proper design and production should be considered in the current Nanoscience and Nanotechnology in order to compute realistic, including defects, macroscopic strengths. The role of defects is expected to be tremendous in carbon nanotube bundles (Pugno, 2006a, 2007a,b). For example, the Silver bridge was an eye-bar chain suspension bridge built in West Virginia in 1928. At the end of the year 1967, it collapsed resulting in the deaths of 46 people. The collapse occurred with the failure of the single eye-bar in a suspension chain, due to a single small defect only 2.54 mm deep. Clearly, in order to avoid this kind of mistakes (see Petroski, 1994, for case histories of error and judgment in engineering), a proper fracture mechanics approach has to be considered.

Dynamic instabilities must also be taken into account, especially for long spans. Self-excited oscillations due to vortexes formation or periodic variation of the aerodynamic lift/moment both depending on the flexural-torsional vibration of the deck, i.e. "flutter", could be extremely dangerous. For example, the Tacoma Narrows bridge, Washington State, was opened on July 1, 1940, and became famous four months later for a dramatic wind-induced structural collapse that was caught on color motion picture film (see http://www.youtube.com/watch?v=3mclp9QmCGs).

In this paper the new concept of "super-bridges", i.e. kilometre-long bridges suspended over carbon nanotube cables, is introduced. The analysis shows that the use of realistic (thus defective) carbon nanotube bundles as suspension cables can enlarge the current limit main span by a factor of ~3 (theoretically by a factor of ~10). We thus suggest that the design of the Messina bridge in Italy, that would require a main span of ~3.3 Km, could benefit by the use of carbon nanotube bundles.

## 2. Strength of carbon nanotube cables

To evaluate the strength of carbon nanotube cables, the $SE^3$ algorithm, formerly proposed by Pugno (2006a), has been adopted (Pugno, Bosia, Carpinteri, 2008). Multiscale simulations are necessary in order to tackle the size scales involved, spanning over ~10 orders of magnitude from nanotube length (~100nm) to kilometre-long cables, and also to provide useful information about cable scaling properties with length. Details are given elsewhere (Pugno, Bosia, Carpinteri, 2008).

The cable is modelled as an ensemble of stochastic "springs", arranged in parallel sections. Linearly increasing strains are applied to the fibre bundle, and at each algorithm iteration the number of fractured springs is computed (fracture occurs when local stress exceeds the nanotube failure strength) and the strain is uniformly redistributed among the remaining intact springs in each section.

In-silico stress-strain experiments have been carried out according to the following hierarchical architecture. Level 1: the nanotubes (single springs, Level 0) are considered with a given elastic modulus and failure strength distribution and composing a 40×1000 lattice or fibre. Level 2: again a 40×1000 lattice composed by second level "springs", each of them identical to the entire fibre analysed at the first level, is analysed with in input the elastic modulus and stochastic strength distribution derived as the output of the numerous simulations to be carried out at the first level. And so on. Four hierarchical levels are sufficient to reach the size-scale of the kilometre from that of the nanometre; only one additional hierarchical level would be sufficient to consider a 100,000 Km-long megacable, as that required in the space elevator design (see Pugno, 2006, 2007a,b for details).

The level 1 simulation is carried out with springs $L_0=10^{-7}$m in length, $w_0=10^{-9}$m in width, with Young's modulus $E_0=10^{12}$Pa and strength $\sigma_f$ randomly distributed according to the nanoscale Weibull statistics (Pugno and Ruoff, 2006) $P(\sigma_f)=1-\exp\left[-\left(\sigma_f/\sigma_0\right)^m\right]$, where $P$ is the cumulative probability. Fitting to experiments (Yu et al., 2000), we have derived for carbon nanotubes $\sigma_0$=34GPa and $m$=2.7 (Pugno and Ruoff, 2006). Then the level 2 is computed, and so on. The results are summarized in Figure 1, in which a strong size-effect is observed.

Given the decaying $\sigma_f$ vs. cable length $L$ obtained from simulations, it is interesting to fit the behaviour with simple analytical scaling laws. Various exist in the literature, and one of the most well-known is the Multi-Fractal Scaling Law (MFSL) proposed by Carpinteri (1994), see also our related commentary (Carpinteri and Pugno, 2005). This law has been recently extended towards the nanoscale (Pugno, 2006b):

$$\frac{\sigma_f}{\sigma_{macro}} = \sqrt{1 + \frac{l_{ch}}{L + l_0}} \tag{1}$$

where $\sigma_f$ is the failure stress, $\sigma_{macro}$ is the macrostrength, $L$ is the structural characteristic size, $l_{ch}$ is a characteristic internal length and $l_0$ is defined via $\sigma_f(l=0) = \sigma_{macro}\sqrt{1 + \frac{l_{ch}}{l_0}} \equiv \sigma_{nano}$, where $\sigma_{nano}$ is the nanostrength. Note that for $l_0 = 0$ this law is identical to the well-known Carpinteri' scaling law (1994). Here, we can choose $\sigma_{nano}$ as the nanotube stochastic strength, i.e. $\sigma_{nano}$=34GPa. The computed macrostrength is $\sigma_{macro}$=10.20GPa. The fit with eq. (1) is shown in Figure 1 ("MFSL cut at the nanoscale"), for the various $L$ considered at the different hierarchical levels (and compared with the classical "MFSL"). The best fit is obtained for $l_{ch}$= 5x10$^{-5}$m, where the analytical law is practically coincident with the simulated results. Thus, for our kilometre-long carbon nanotube cables we can assume a plausible strength $\sigma_C = \sigma_{macro} \approx 10\text{GPa}$. We must note that 10GPa-strong carbon nanotube fibers are today available (Koziol et al., 2007), suggesting that long cables with a similar strength could be realized in the near future.

## 3. Bridge limit main span

Let us consider a classical suspended-deck bridge, as that related to the existing longest bridge (Akashi-Kaikyo bridge, with a main span of 1991 meters) and reported in Figure 2. It is well known that neglecting the cable weight with respect to the bridge weight leads to a parabolic shape of the cable. In particular, indicating with $q$ the weight per unit length of the bridge and with $T_O$ the horizontal component of the tension at the towers (see Figure 2), the shape equation for the cable, reported in all the text books of structural mechanics (e.g., see Carpinteri, 1997), is $\mathrm{d}^2 y/\mathrm{d}z^2 = -q/T_O$. The weight per unit length is $q \approx \gamma A$, where $\gamma$ is the specific weight of the bridge (increased in order to take into account the weight of the accidental loads, e.g. vehicles, and the equivalent weight of the cables; it is expected to be strongly reduced using carbon nanotube technology, thus a parabola rather than a catenary is the expected final shape) and $A$ is its cross-section area. Noting that $\mathrm{d}y/\mathrm{d}z(z = L/2) = y(z=0) = 0$, the cable shape $y(z) = \frac{qL^2}{2T_O}\left(\frac{z}{L} - \frac{z^2}{L^2}\right)$ is deduced by a trivial integration. Let us indicate with $h = y(z = L/2) = qL^2/(8T_O)$ the cable "height" and with $\tan\alpha = \mathrm{d}y/\mathrm{d}z(z=0) = qL/(2T_0) = 4h/L$ the cable slope at the towers. Their compression is $T_V = qL/2$, whereas the tension in the cable is $T = T_V/\sin\alpha$. Evidently, the maximum tension of the cable is $T = T_C = \sigma_C A_C$, where $\sigma_C$ is the cable strength and $A_C$ is its cross-sectional area. Rearranging the previous equations we find the limit main span $L_{\lim}$ according to the following formula:

$$\frac{L_{\lim}}{h} = \sqrt{8\left(\sqrt{1 + \frac{\sigma_C^2 A_C^2}{\gamma^2 A^2 h^2}} - 1\right)} \tag{2}$$

Note that, for $h/L \to 0$, $L_{\lim} \to \sqrt{8h\sigma_C A_C/(\gamma A)}$, in contrast with the case of a stayed bridge for which $L_{\lim} \approx \sqrt{8h\sigma_B/\gamma}$ (see Ryall, Parke, Harding, 2000) where $\sigma_B$ is the strength of the material composing the deck. As anticipated, the difference is due to the fact that the main stress fields in a suspended-deck bridge are tension of the cables and compression of the towers, whereas in the stayed bridges is compression of the towers and the deck. This confirms that the use of carbon nanotube strands is not very effective, suggesting that a suspended-deck rather than a stayed structure is required in the design of the super-bridge, at least with respect to static considerations.

Fatigue of the cables, could also play a role. The limit span would be again given by eq. (2), interpreting $\sigma_C$ as the fatigue limit. Thus the same strong gain by using carbon nanotube cables is expected.

In addition to the strength main requirement, we must limit the maximum vertical displacement of the bridge; it takes place at the middle section of the main span. Geometrically we have calculated the length *l* of the parabolic cable as:

$$\frac{l}{L} = \frac{\sqrt{16h^2/L^2+1}}{2} + \frac{L\ln\left(4h/L+\sqrt{16h^2/L^2+1}\right)}{8h} \tag{3}$$

Assuming a sufficiently high density of the vertical cables, connecting the main cable with the deck, their compliance can be neglected; furthermore, to be conservative, we assume a vanishing stiffness of the deck. If a cable additional strain $\Delta\varepsilon_C$ is imposed (e.g. by accidental loads) the length of the parabolic cable becomes $l' = l(1+\Delta\varepsilon_C)$; introducing this value into eq. (3) yields the new cable height $h'$ and thus the bridge maximum deflection $\eta$, according to a finite kinematics nonlinear approach, as:

$$\frac{\eta}{L} = \frac{h'(l') - h(l)}{L} \tag{4}$$

For example, considering $h/L = 0.1$, from eq. (3) or Table 2 we find $l/L = 1.026061$; and assuming $\Delta\varepsilon_C = 10^{-3}$ ($\varepsilon_C \approx 10^{-2}$) we have $l'/L = 1.027087$ and, again from eq. (3) or Table 2, we deduce $h'/L = 0.102$, thus $\eta/L = 0.002$, i.e. $2\,\mathrm{m/Km}$ is the expected maximum deflection. It could be tolerated by a proper design of the bridge.

Dynamic instabilities could also play a key role in super-bridges (see Ryall, Parke, Harding, 2000). The most important design requirements for the deck structure are low weight, high torsional stiffness and good aerodynamics characteristics. The effects of wind on traffic has also to be considered. The main aerodynamics actions and effects are the mean wind loading (drag force, proportional to $C_D V^2$, where $V$ is the wind velocity and $C_D$ is the drag coefficient; it is today around 0.075, compared to typical values of about 0.25), the randomly fluctuating turbulent components of the wind (buffeting force, proportional to $V^\alpha$ with $\alpha > 2$), the vortex shedding (rarely giving rise to problem and governed by the Strouhal number $N_S = fW/V$, where $W$ is the width of the bridge and $f$ is the frequency of vortex formation, taking place at $N_S \approx 0.1$) and the aerodynamic instabilities, in transverse bending (galloping), torsion (stall flutter) or coupled torsion and bending (classical flutter). A detailed analysis is out of the scope of the present paper; however, we may note that approximated formulas exist, e.g. the Selberg critical speed for classical flutter (see Ryall, Parke, Harding, 2000 and related references for details):

$$V_C \approx 4 f_T W (1 - f_B / f_T) \sqrt{\frac{\gamma A I}{\rho W^3}} \tag{5}$$

where $f_{B,T}$ are the bending/torsional natural frequencies, $I$ is the polar moment of inertia of the bridge cross-section and $\rho$ is the air density.

Since $f_{B,T} \propto L^{-2}\sqrt{EI/(\gamma A)}$, super-bridges are clearly critical with respect to dynamic instabilities, which thus require a further a detailed analysis as well as new proposals for technical solutions. Roughly speaking, to avoid resonances, aerodynamic cross-sections and bridge stiffening, to increase the fundamental frequencies, have to be considered. To increase the bridge stiffness larger frames could be used and the introduction of strands, as in a stayed bridge, could be added in parallel to the main suspension cable. In order to avoid their monolateral effect, strands also below the deck could be introduced, considering towers protruding also downwards. Lateral strands could also increase the out-of-plane stiffness of the bridge, leading to a super-bridge anchored similarly to a spider cobweb.

As an example of application of eq. (2), we have computed the limit main spans (according to static considerations) of the current existing 19 suspended-deck bridges longer than one kilometre, see Table 1, assuming to have substituted their cables with carbon nanotube bundles (thus maintaining the same deck and cables cross-sections) finding spans of up to ~6.3 Km, (information from http://en.wikipedia.org/wiki/Suspension_bridge). We have assumed here $L >> h$, i.e. $L_{\lim} \approx \sqrt{8h\sigma_C A_C/(\gamma A)}$ thus predicting a gain by a factor of $\sqrt{10}$, in order to avoid to know the details of the existing bridges, but more precise calculations are scientifically trivial.

## 4. Statistical data analysis

The plausibility of these lengths is confirmed by a statistical analysis on the existing 100 longest suspended bridges, Figure 3. We have found a Moore-like (i.e. exponential) law for the "records" (considering the bridges having a span following a positive monotonic trend with time, i.e. described by the points belonging to the upper enveloping curve of the data reported in Figure 3), in which the doubling of the capacity (here the main span) per year is substituted by the factor $e^{0.0137} \approx 1.01379$, Figure 4: i.e., the length($L$[m])-year($Y$) relationship is given by $L \propto 1.014^Y$. This corresponds to the prediction for a "conventional" bridge realization according to:

$$Y \approx \frac{\ln\left[L[\mathrm{m}]/\left(2\times 10^{-9}\right)\right]}{0.0137} \tag{6}$$

resulting for the Messina bridge in 2053. Even considering only the most recent three data in Fig. 4 would lead to the realization in 2050 ($L \propto 1.01267^Y$). In spite of this, if the technological revolution, expected thanks to the introduction of carbon nanotube bundles, is considered, an abrupt increment of the order of three in the achievable lengths is expected. An event similar, from both a qualitative and quantitative point of view, happened after the USA crisis around the year 1929, as can be easily observed in Fig. 4, perhaps mainly caused by social rather than technological innovations (the introduction of steel dated back to 1883 with the Brooklyn's bridge). In addition, we may note that the longest existing suspended railways bridge has a length of only ~1.1Km, thus one half of the longest bridge. This is a consequence of the larger stiffness required by railways. The Messina bridge, which has to include the railways, needs a length that is three times larger than that of the longest railways suspended bridge. We have shown that the factor of three can be gained, in a near future, thanks to carbon nanotube cables. Thus, we believe that the use of carbon nanotube bundles is becoming the sole feasible solution for the realization in the near future of the Messina bridge.

By the statistical data treatment, we have also found that the occurrence frequencies of the analyzed 100 bridges follow a Zipf's law, Figure 5. It predicts that, out of a given population, the frequency of elements of rank $k$ is $f \propto k^{-s}$, where $s$ is the exponent of the distribution. For the considered data set we find ($\mathrm{R}^2 = 0.8363$):

$$f \approx 39.026 k^{-1.1615} \tag{7}$$

Note that $s \approx 1.1615$ is close to the unitary value of the well-known "1/*f* function".

A similar, simplified but more intuitive, analysis could be directly carried out considering the main span instead of the rank; we find $f \approx 3 \times 10^8 L^{-2.5287}$, see Figure 6.

## 5. Conclusions

In this paper, the new concept of "super-bridges", i.e. kilometre-long bridges suspended over carbon nanotube cables, has been introduced. The analysis has shown that the use of realistic (thus defective, see eq. (1)) carbon nanotube bundles, expected to be one order of magnitude stronger than the current ones, as suspension cables can enlarge the current limit main span by a factor of ~3, (see eq. (2)). Computing the limit main spans of the current existing 19 suspended-deck bridges longer than one kilometre and assuming to have substituted their cables with carbon nanotube bundles (maintaining the same deck and cables cross-sections) we have found main spans in the 3−6 kilometre range. Bridge deflection can be easily limited (see eqs. (3) and (4)). Even if dynamic self-excited resonances have not been studied in details here, we believe that flutter (see eq. (5)) could be avoided by additional strands, rendering the super-bridge anchored as a spider cobweb. We thus suggest that the design of the Messina bridge in Italy, that would require a main span of ~3.3 Km, could benefit by the use of carbon nanotube bundles.

We have found a Moore-like (i.e. exponential) law for the "records" (see eq. (6)) corresponding to the prediction for a "conventional" realization of the Messina bridge around the year 2050. In addition, we have noted that the Messina bridge, which has to include the railways, needs a length that is three times larger than that of the existing longest railways suspended bridge. We have shown that the factor of three can be gained thanks to carbon nanotube cables. By the statistical data treatment, we have also found that the occurrence frequencies of the analyzed 100 bridges follow a Zipf's law, (see eq. (7)).

Concluding, the analysis shows that the use of carbon nanotube bundles is becoming the sole feasible solution for the realization in our days of super-bridges, as those required across the Straits of Bab al Mandab, Messina or Gibraltar (main spans ~ 2.7, 3.3 or 3.5Km, respectively); the first and last ones would be intercontinental bridges, between the Arabian Peninsula and the Horn of Africa (across the Red Sea) or between Spain and Morocco (connecting the Atlantic Ocean to the Mediterranean Sea), respectively.

We must note that 10GPa-strong carbon nanotube fibers are today available (Koziol et al., 2007), suggesting that long cables with a similar strength could be realized in the near future. Regarding the cost analysis, we may note that carbon nanotubes had in 2006 an approximate price of \$25/gram. Over the past two years, scale up of multi-wall carbon nanotube production has led to a dramatic price decrease (Arkema, Bayer Material Sciences, Showa Denko), down to \$150/kg for semi-industrial applications. The run for industrial carbon nanotube production plants has started in order to achieve a sustainable business with the commercialization of these high-tech materials with a mid-term price target of \$45/kg (http://www.electronics.ca/presscenter/articles/743/1/Carbon-Nanotube-Production-Dramatic-Price-Decrease-Down-to-150kg-for-Semi-Industrial-applications/Page1.html). Accordingly, the cost of a 10Km-length carbon nanotube cable with a diameter of 1m is expected to be of the order of 1 billion dollars, i.e. ~1/10 of the characteristic cost of a kilometre-long suspended bridge, suggesting that our solution is also economically feasible.

The authors are supported by the ‘‘Bando Ricerca Scientifica Piemonte 2006’’ – BIADS: Novel biomaterials for intraoperative adjustable devices for fine tuning of prostheses shape and performance in surgery, and thank Ing. Matteo Accardi for commenting the manuscript.

**FIGURES**

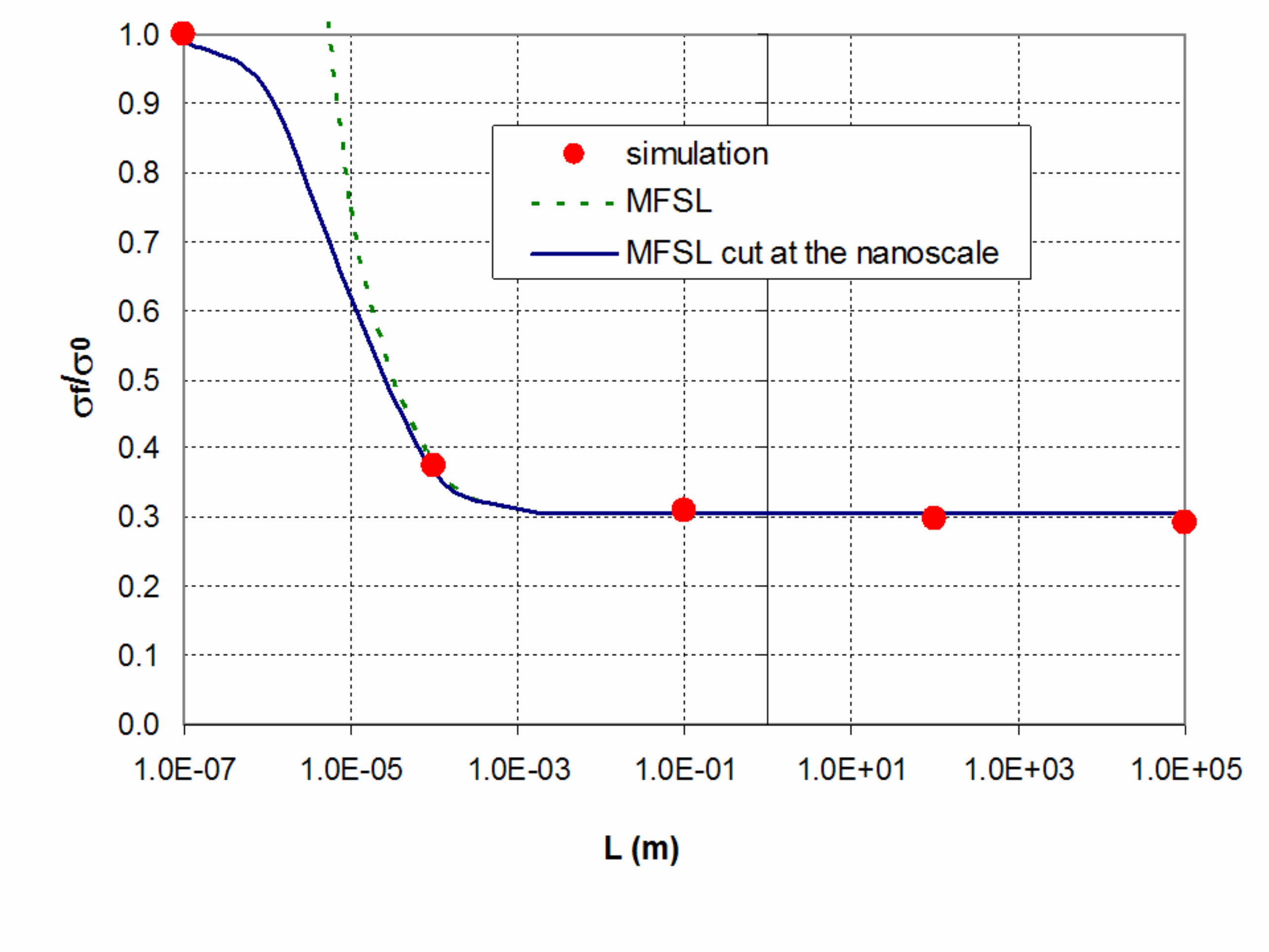

Figure 1: Comparison between simulations and analytical laws (see text) for the failure strength of the nanotube bundle as a function of its length; the asymptote is at 10.20GPa (figure adapted from Pugno, Bosia and Carpinteri, 2008).

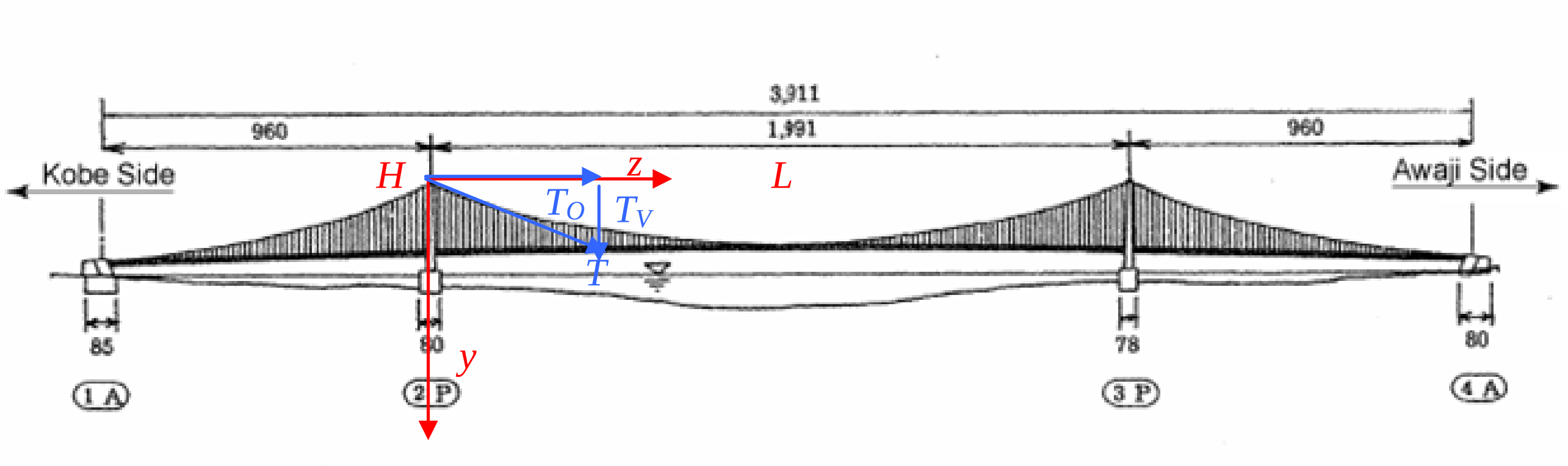

Figure 2: The structural scheme of the Akashi-Kaikyo bridge, with a main span $L$ of 1991 meters (or 3911 feet) suspended over a parabolic cable with geometry $y(z)$, between two towers of height $H$.

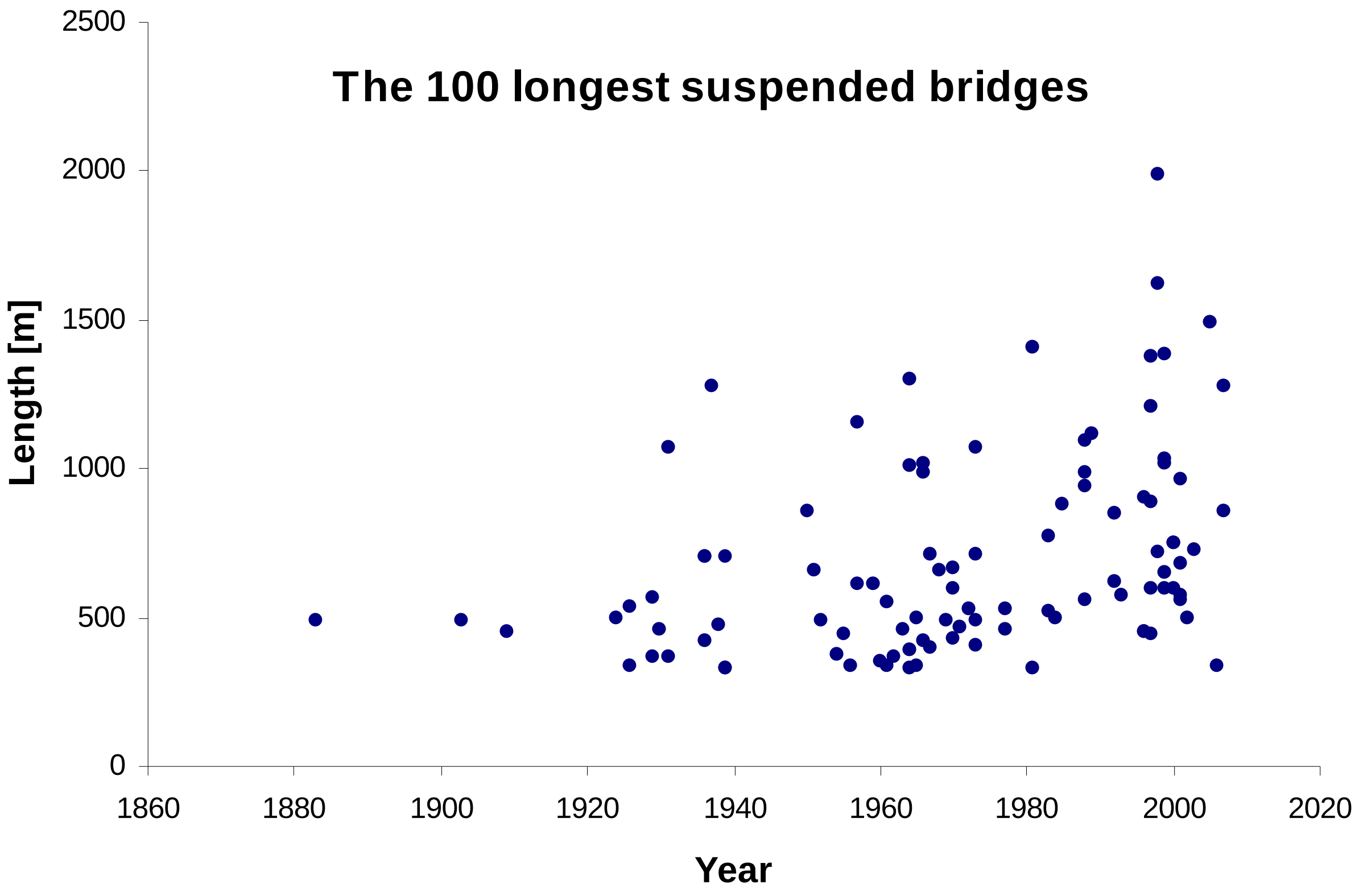

Figure 3: The 100 longest suspended bridges; the first 19 bridges, longer than 1Km, are reported in Table 1. The oldest considered bridge is the Brooklyn's (1883), the first with steel cables; before it, two notable bridges with main spans of 308 and 322 meters were opened in 1849 and 1867, respectively.

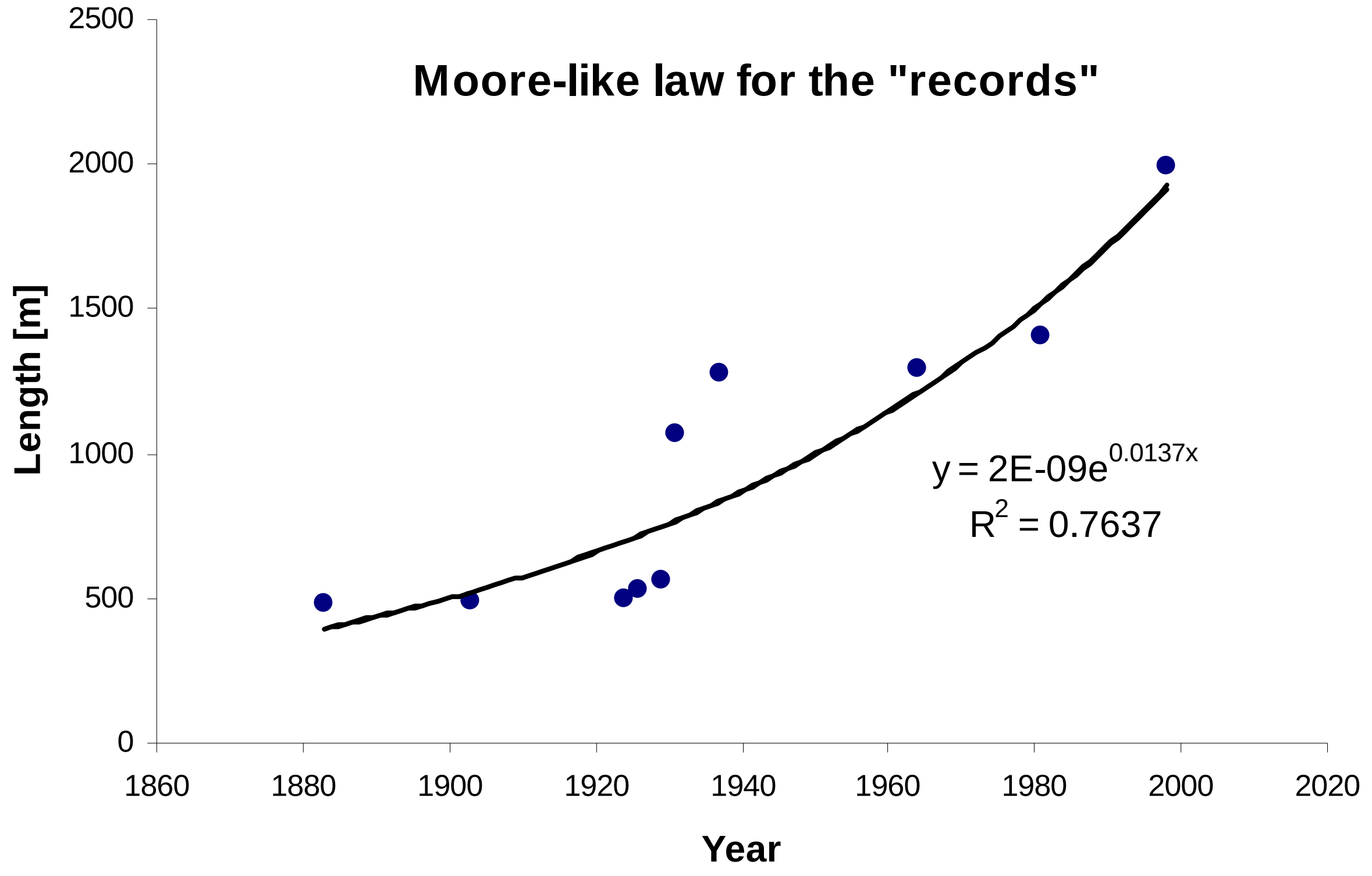

Figure 4: Moore-like law for long suspended bridges. The considered bridges here are only the "records".

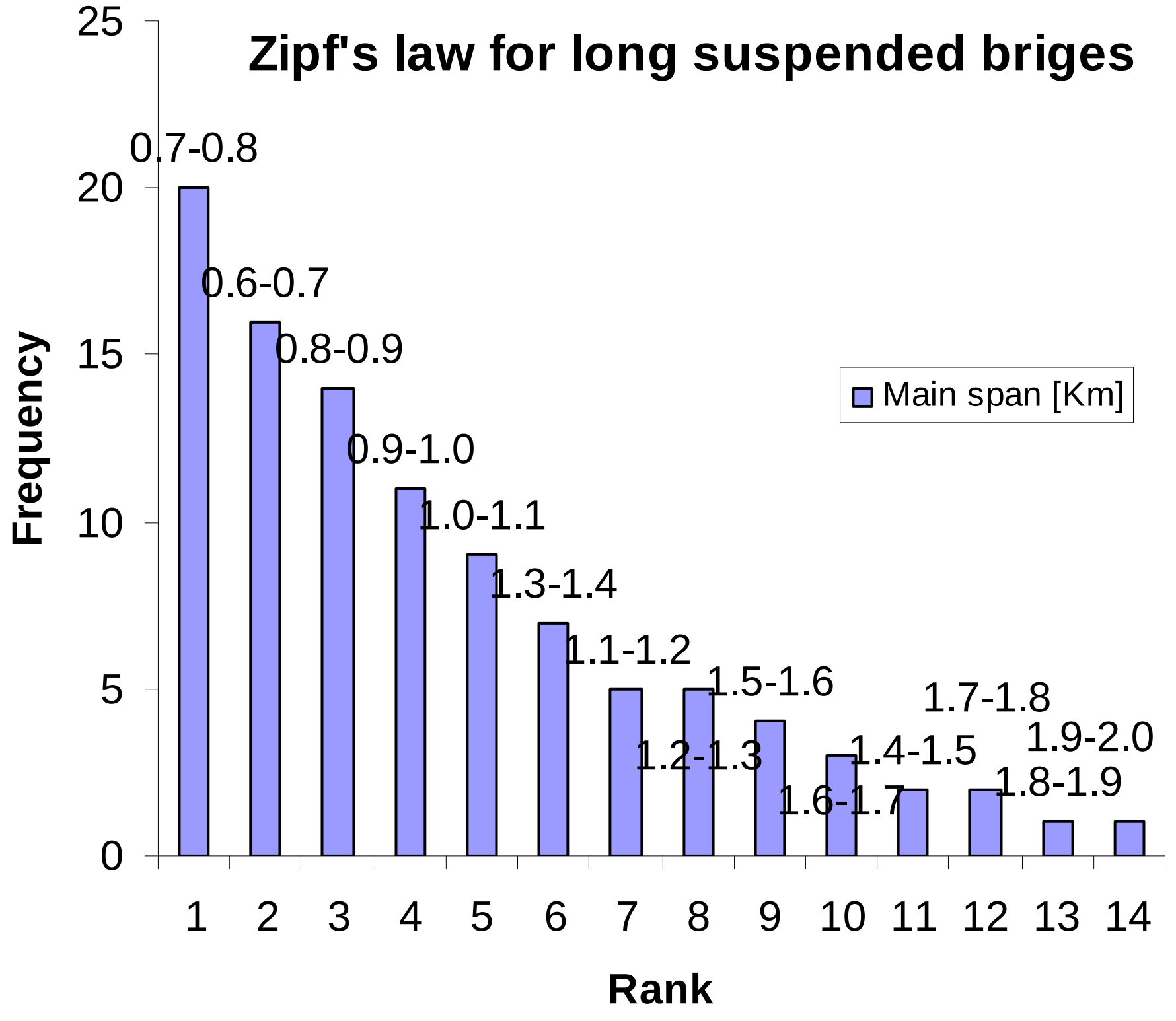

Figure 5: Zipf's law for long suspended bridges. The frequencies are evaluated considering classes of main spans with increment equal to 100m. Note that the sum of the frequencies is 100, that is the number of the analyzed bridges.

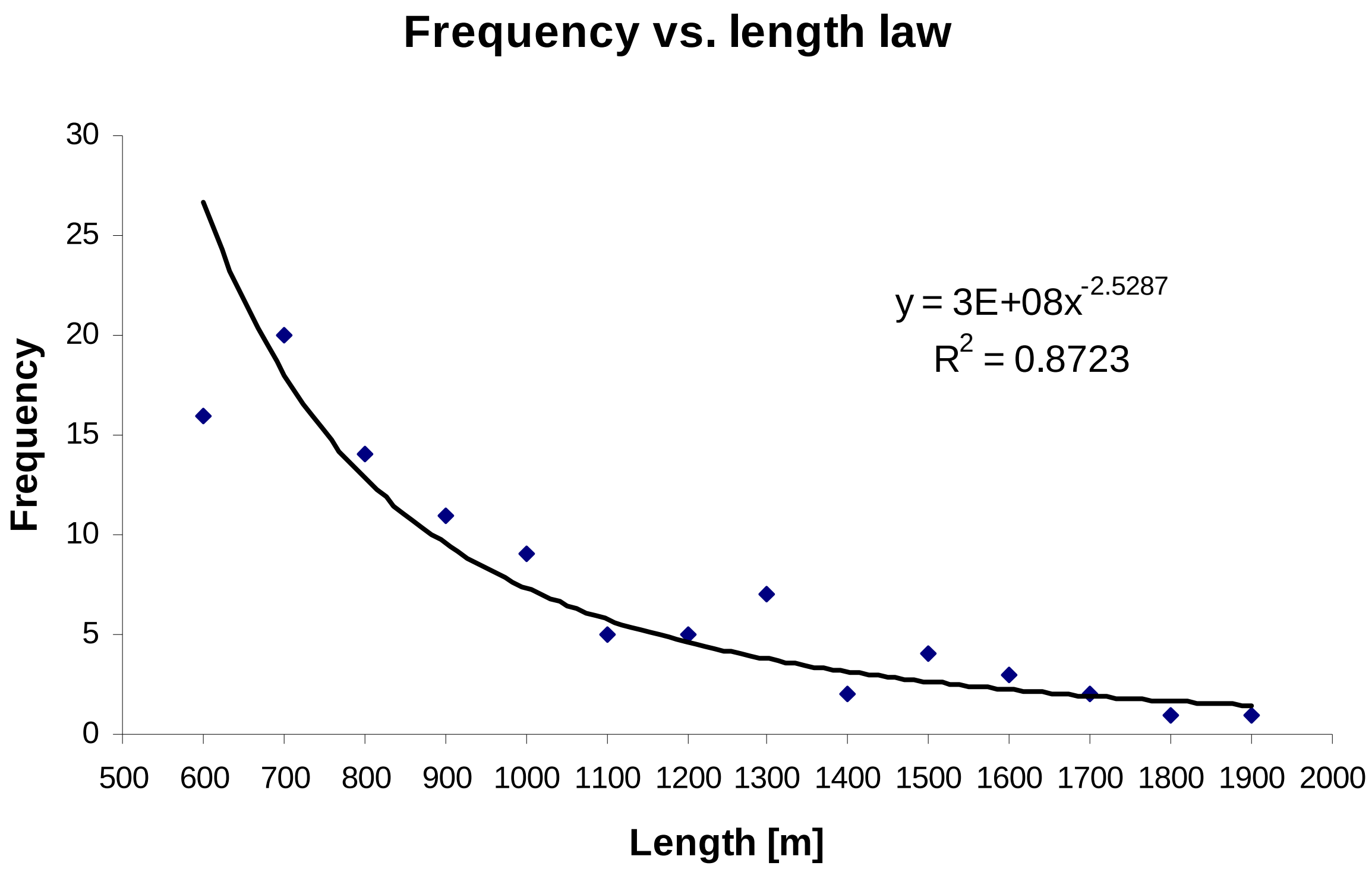

Figure 6: Frequency vs. length for long suspended bridges. The frequencies are evaluated considering classes of main spans with increment equal to 100m. Note that the sum of the frequencies is 100, that is the number of the analyzed bridges.

# TABLES

| Name | Structure | Year opened | Place | Bridge main span [meters] |
|---|---|---|---|---|
| Akashi-Kaikyo | 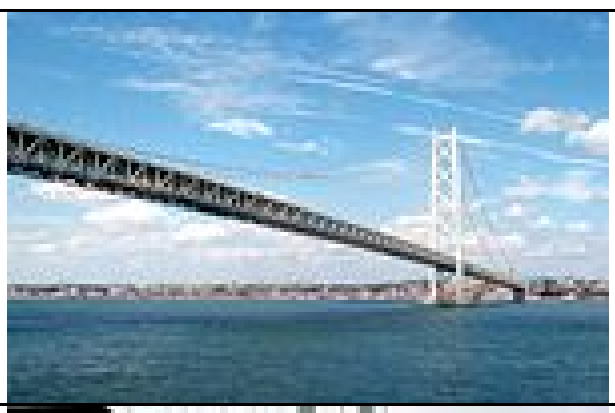 | 1998 | Kobe-Naruto Route, Japan | 1991 |
| Great Belt | 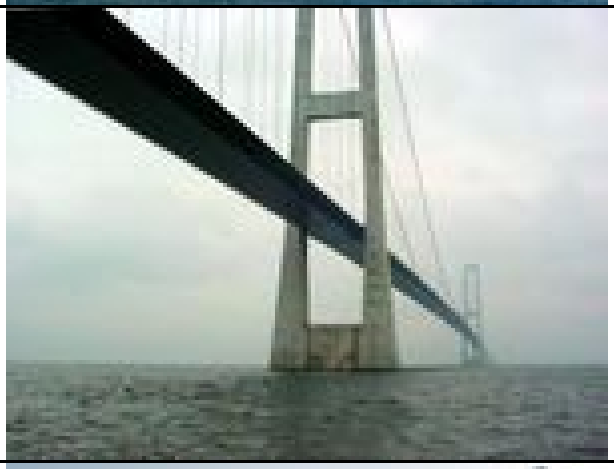 | 1998 | Halsskov-Sprogø, Denmark | 1624 |
| Runyang | 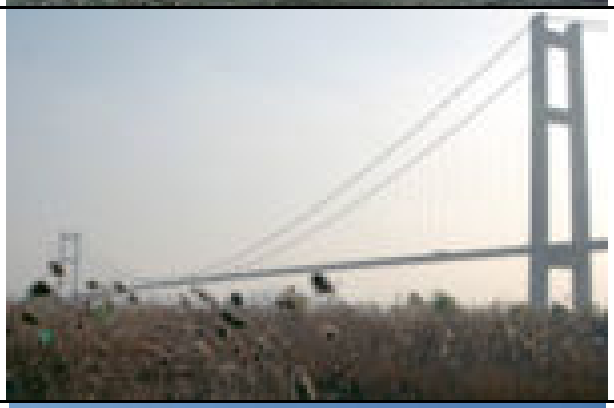 | 2005 | Yangtze River, China | 1490 |
| Humber | 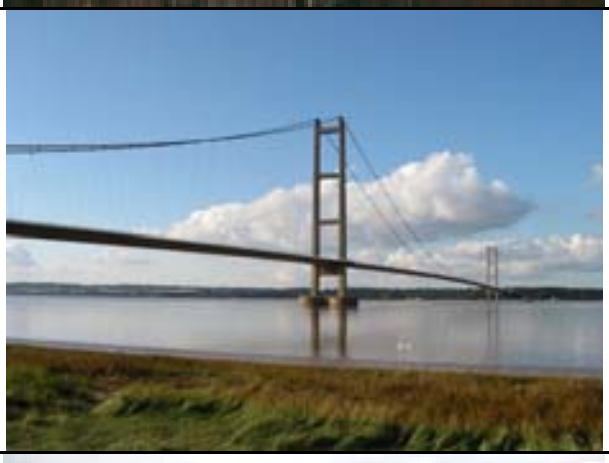 | 1981 | Barton-upon-Humber - Kingston upon Hull, United Kingdom | 1410 |
| Jiangyin Suspension | 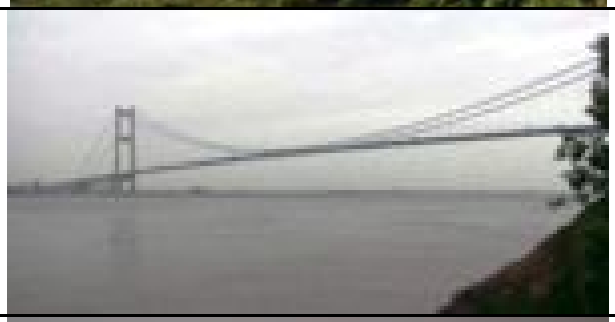 | 1999 | Yangtze River, China | 1385 |
| Tsing Ma | 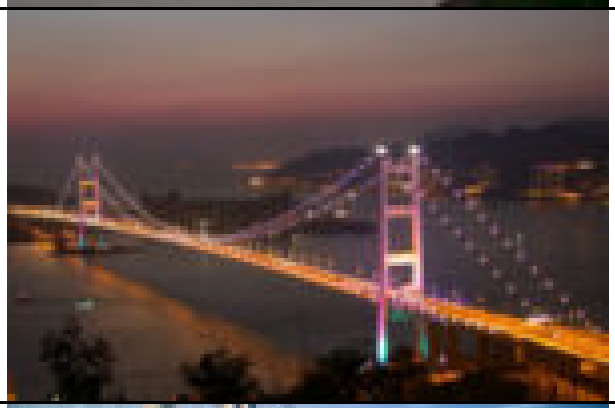 | 1997 | Tsing Yi-Ma Wan, Hong Kong, China | 1377 |
| Verrazano-Narrows | 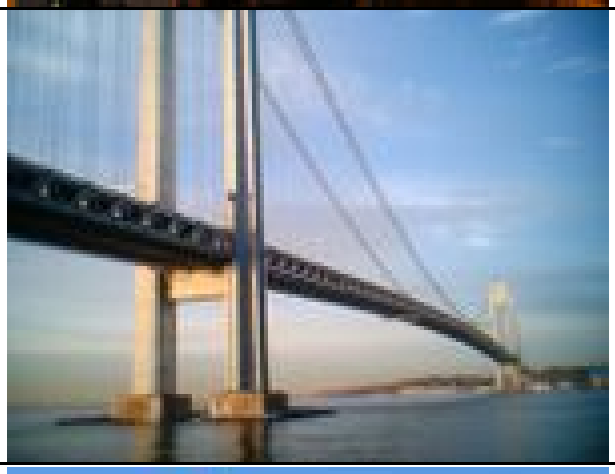 | 1964 | New York City, Brooklyn–Staten Island, USA | 1298 |
| Golden Gate | 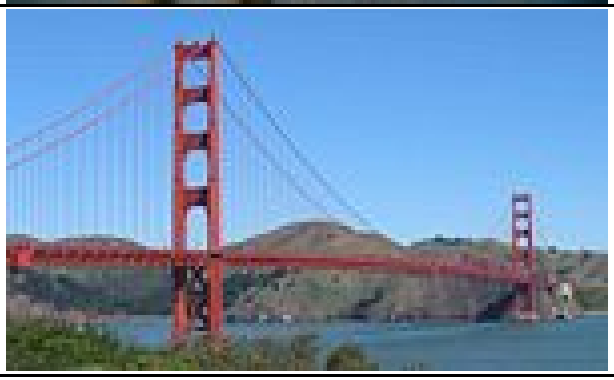 | 1937 | San Francisco-Marin County, CA, USA | 1280 |
| Yangluo | 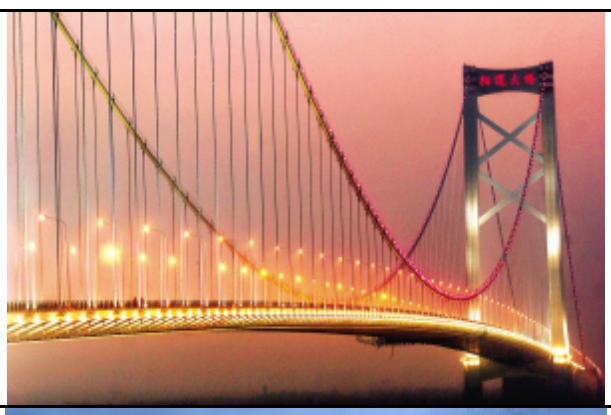 | 2007 | Yangtze River, China | 1280 |
| Högakustenbron | 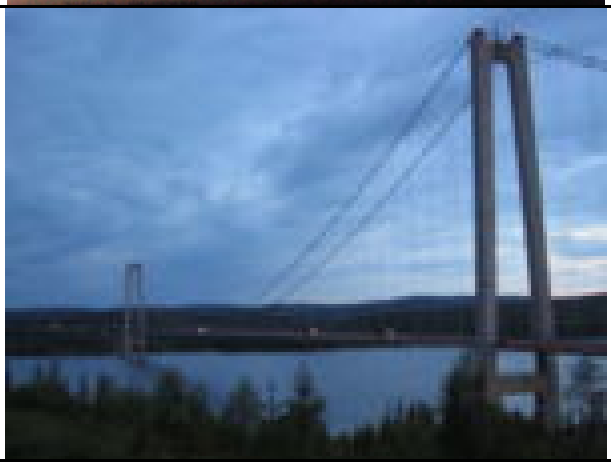 | 1997 | Ångermanälven river, Sweden | 1210 |
| Mackinac | 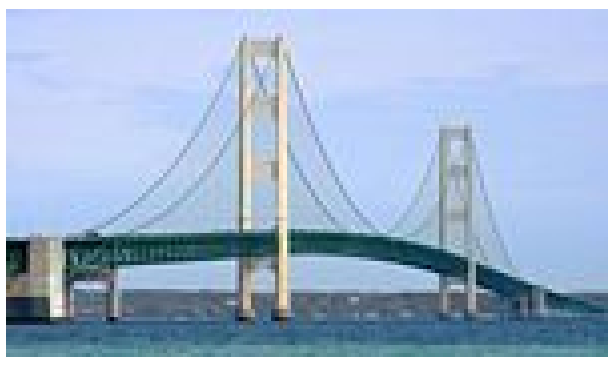 | 1957 | Mackinaw City - St. Ignace, Michigan USA | 1158 |
| Minami Bisan-Seto | 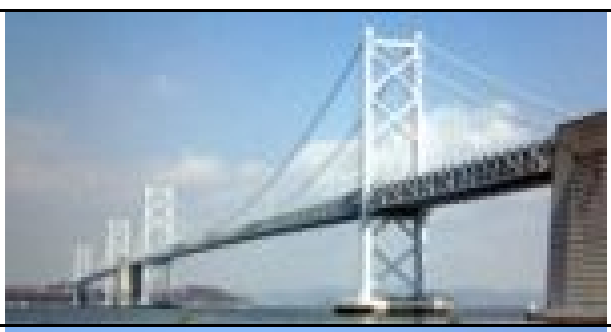 | 1989 | Kojima-Sakaide Route, Japan | 1118 |
| Fatih Sultan Mehmet | 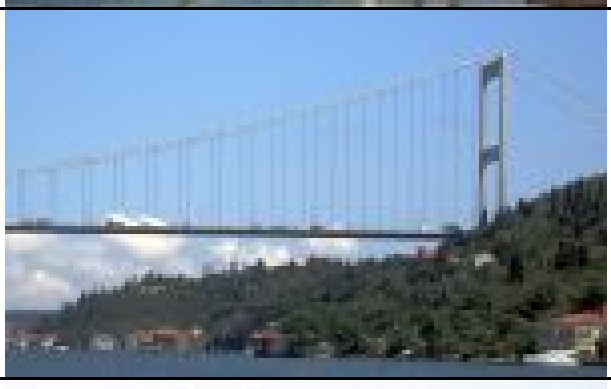 | 1988 | Istanbul, Turkey | 1090 |
| Boğaziçi | 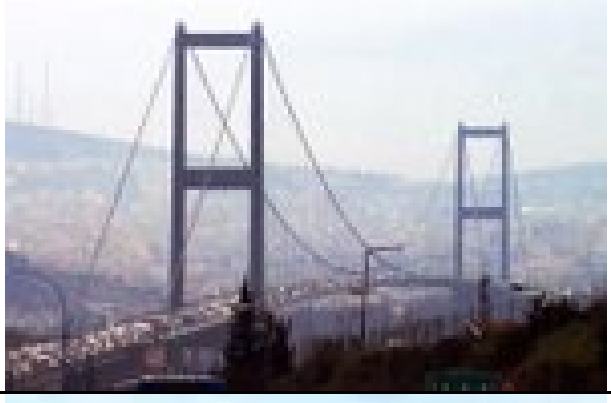 | 1973 | Istanbul, Turkey | 1074 |
| Geoge Washington | 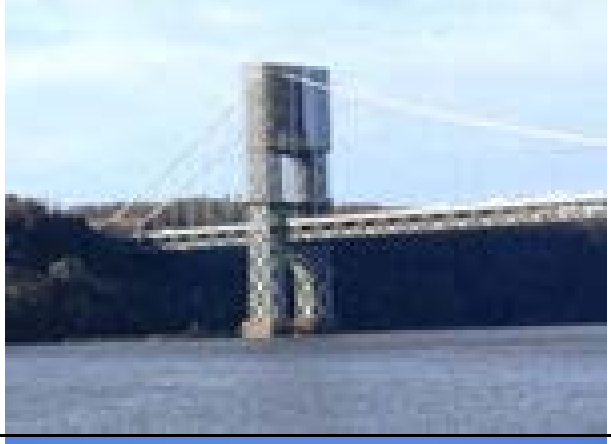 | 1931 | Fort Lee, NJ - New York, NY, USA | 1067 |
| Third Kurushima-Kaikyo | 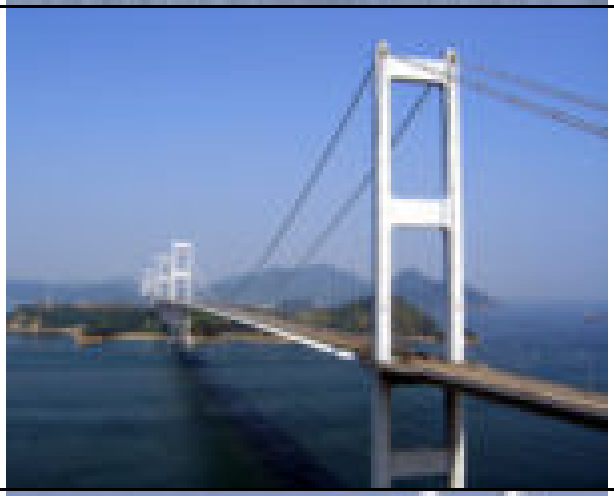 | 1999 | Onomichi-Imabari Route, Japan | 1030 |
| Second Kurushima-Kaikyo | 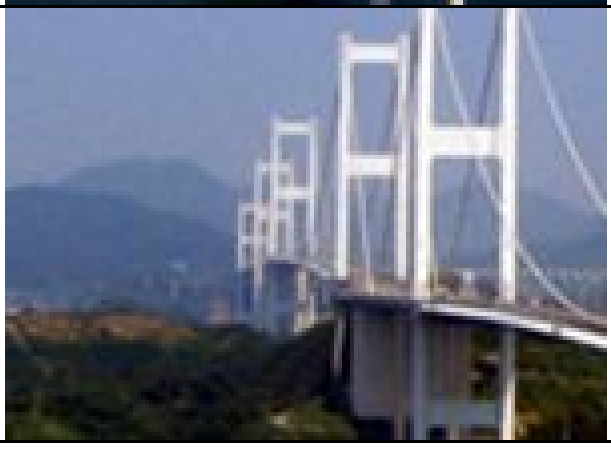 | 1999 | Onomichi-Imabari Route, Japan | 1020 |
| Ponte 25 de Abril | 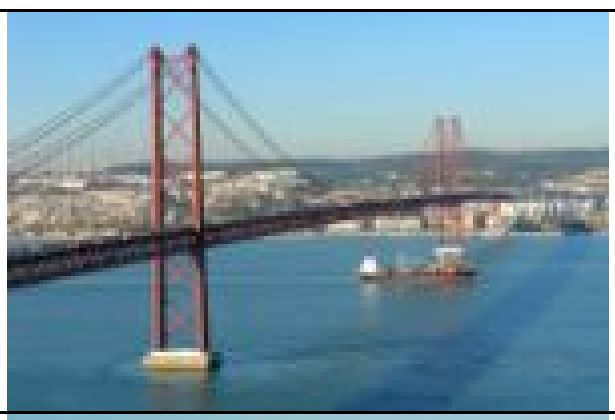 | 1966 | Lisbon, Portugal | 1013 |
| Forth Road | 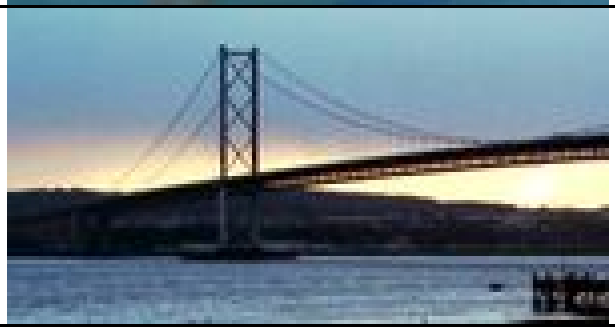 | 1964 | Firth of Forth, United Kingdom | 1006 |

Table 1: The current bridges longer than 1Km (adapted from http://en.wikipedia.org/wiki/Suspension_bridge). The expected achievable main spans using carbon nanotube bundles with the same cross-section of the existing cables are approximately 3 times. We have considered for the existing high strength (0.65-0.70GPa, see Ryall, Parke, Harding, 2000) cables an upper limit of 1GPa.

| $h/L$ | $l/L$ |
|---|---|
| 0.08 | 1.016814 |
| 0.081 | 1.01723 |
| 0.082 | 1.017652 |
| 0.083 | 1.018078 |
| 0.084 | 1.01851 |
| 0.085 | 1.018946 |
| 0.086 | 1.019387 |
| 0.087 | 1.019832 |
| 0.088 | 1.020283 |
| 0.089 | 1.020738 |
| 0.09 | 1.021198 |
| 0.091 | 1.021663 |
| 0.092 | 1.022133 |
| 0.093 | 1.022607 |
| 0.094 | 1.023087 |
| 0.095 | 1.023571 |
| 0.096 | 1.024059 |
| 0.097 | 1.024552 |
| 0.098 | 1.025051 |
| 0.099 | 1.025553 |
| 0.1 | 1.026061 |
| 0.101 | 1.026573 |
| 0.102 | 1.027089 |
| 0.103 | 1.027611 |
| 0.104 | 1.028137 |
| 0.105 | 1.028667 |
| 0.106 | 1.029202 |
| 0.107 | 1.029742 |
| 0.108 | 1.030287 |
| 0.109 | 1.030835 |
| 0.11 | 1.031389 |
| 0.111 | 1.031947 |
| 0.112 | 1.032509 |
| 0.113 | 1.033077 |
| 0.114 | 1.033648 |
| 0.115 | 1.034224 |
| 0.116 | 1.034805 |
| 0.117 | 1.03539 |
| 0.118 | 1.035979 |
| 0.119 | 1.036573 |
| 0.12 | 1.037171 |

Table 2: Bridge deflection, eq. (3). For example, plausibly considering $h/L = 0.1$ gives $l/L = 1.026061$; assuming $\Delta\varepsilon_C = 10^{-3}$ ($\varepsilon_C \approx 10^{-2}$) we have $l^{/}/L = 1.027087$ corresponding to $h^{/}/L = 0.102$, thus $\eta/L = 0.002$, i.e. $2\mathrm{m}/\mathrm{Km}$ is the expected maximum deflection.